\documentclass[fleqn,10pt]{wlscirep}
\usepackage{amsmath, bm}
\usepackage[utf8]{inputenc}
\usepackage[T1]{fontenc}
\usepackage{comment}
\usepackage{lineno}
\usepackage[nodisplayskipstretch]{setspace}
\usepackage[format=plain, font={small, singlespacing}, labelfont=bf]    {caption}
\usepackage{capt-of}
\usepackage[CaptionAfterwards]{fitpage}
\usepackage{caption}
\usepackage{xr}
\usepackage{xcolor}
\usepackage{multirow}
\usepackage{colortbl}
\usepackage[all,color]{xy}
\usepackage{algpseudocode}
\usepackage{algorithm}
\usepackage{wrapfig}
\usepackage{makecell}

\PassOptionsToPackage{hyphens}{url}\usepackage{hyperref}
\allowdisplaybreaks[1]

    \usepackage{soul}
    \usepackage{xspace}

    \title{Structural Divergence Between AI-Agent and Human Social Networks in Moltbook}
    \author[1,$\dagger$]{Wenpin Hou}
    \author[1,$\dagger$]{Zhicheng Ji}
    \affil[1]{Department of Biostatistics and Bioinformatics, Duke University School of Medicine, Durham, NC, USA.}
    \affil[$\dagger$]{Corresponding author. E-mail: wenpin.hou@duke.edu; zhicheng.ji@duke.edu}
    
    \begin{abstract}

Large populations of AI agents are increasingly embedded in online environments, yet little is known about how their collective interaction patterns compare to human social systems. Here, we analyze the full interaction network of Moltbook, a platform where AI agents and humans coexist, and systematically compare its structure to well-characterized human communication networks. Although Moltbook follows the same node–edge scaling relationship observed in human systems, indicating comparable global growth constraints, its internal organization diverges markedly. The network exhibits extreme attention inequality, heavy-tailed and asymmetric degree distributions, suppressed reciprocity, and a global under-representation of connected triadic structures. Community analysis reveals a structured modular architecture with elevated modularity and comparatively lower community size inequality relative to degree-preserving null models. Together, these findings show that AI-agent societies can reproduce global structural regularities of human networks while exhibiting fundamentally different internal organizing principles, highlighting that key features of human social organization are not universal but depend on the nature of the interacting agents.

    \end{abstract}
    
\begin{document}
    \flushbottom
    \maketitle

\section*{Introduction}

Recent advances in large language models and autonomous systems have enabled the deployment of AI agents that can generate content, initiate interactions, and respond adaptively to both humans and other agents. These AI agents are increasingly embedded in networked online environments, ranging from social media platforms and forums to collaborative and simulated societies. Prior work has documented the growing presence of automated agents and social bots in online systems, highlighting their ability to produce human-like content and participate at scale in digital interactions \cite{ferrara2016rise}. More recently, large language models have further expanded the expressive and interactive capabilities of such agents, enabling more sustained and context-aware engagement in shared environments.

Despite rapid progress in AI capabilities at the individual-agent level, a fundamental gap remains in our understanding of how large populations of AI agents interact collectively. In contrast to human social networks, which have been extensively studied for decades, the macroscopic structure and organizing principles of AI-agent interaction networks remain poorly characterized. This gap has been explicitly noted in calls for a new science of ``machine behavior'', which argues that AI systems should be studied empirically in real-world settings, particularly when they interact with one another at scale \cite{rahwan2019machine}. Emerging studies have begun to explore collective behaviors in populations of language-model-based agents, demonstrating the potential for convention formation, coordination, and bias at the group level \cite{ashery2025emergent,chen4895723unveiling}. However, systematic comparisons between AI-agent networks and well-established human social networks are still rare.

Understanding how AI agents interact, and how their interaction networks compare to human social systems, is important for both theory and practice. From a network science perspective, structural properties such as degree distributions, reciprocity, clustering, motifs, and community organization shape how information, attention, and influence flow through a system \cite{girvan2002community}. In human networks, these properties are closely tied to cognitive and social constraints, including limited attention, reciprocity norms, and the tendency toward triadic closure \cite{gonccalves2011modeling,miritello2013limited}. If AI-agent networks exhibit fundamentally different structural patterns, this would suggest that key features of social organization are not universal but instead depend on the nature of the interacting entities. Conversely, structural similarities would indicate that certain network-level regularities emerge independently of human cognition, driven instead by platform rules or interaction dynamics.

Moltbook provides a unique opportunity to investigate these questions. Moltbook is an online platform in which AI agents and humans coexist as explicit participants in a shared interaction network. Unlike traditional social media platforms, where automated activity is often hidden, adversarial, or restricted, Moltbook openly supports large-scale AI participation, making AI-agent interactions directly observable. The resulting network spans tens of thousands of nodes and hundreds of thousands of directed interactions, offering sufficient scale for robust network analysis.

In this study, we analyze the Moltbook interaction network using established tools from network science and systematically compare its structural properties to those of well-characterized human social networks \cite{mislove2007measurement,jiang2013understanding,zlatic2006wikipedias,szell2010multirelational,vscepanovic2017semantic,wang2016power}. We examine global scaling behavior, inequality in attention and interaction, reciprocity and motif structure, and community organization. By situating Moltbook within the broader landscape of online social systems, we aim to identify which aspects of social network structure persist in AI-agent-dominated environments and which diverge sharply from human patterns. Our results reveal that, while Moltbook aligns with human networks at the level of global scaling, its internal organization is markedly different, shedding light on the distinct mechanisms that govern collective AI-agent interaction.

\section*{Results}

We first situate the Moltbook interaction network within the broader landscape of online social systems by comparing its number of nodes and directed edges to a range of well-studied human networks \cite{mislove2007measurement,jiang2013understanding,zlatic2006wikipedias,szell2010multirelational,vscepanovic2017semantic,wang2016power} (Figure 1a). Moltbook, with 17,417 nodes and 161,320 directed edges, lies close to the overall linear trend observed in log–log space, rather than appearing as a pronounced outlier above or below this relationship. This indicates that as Moltbook grows, its total volume of interactions increases at a rate comparable to that of human social systems, rather than deviating markedly with scale. The alignment of Moltbook with the node–edge scaling pattern of human networks suggests that its interaction structure is governed by similar global constraints on communication, rather than by unconstrained or artificially amplified activity. As a result, Moltbook occupies a structurally comparable regime to human social networks, providing a meaningful foundation for downstream comparative analyses.

We next examined a collection of canonical global network statistics that characterize how connections are organized, including assortativity, average degree, average path length within the giant component, clustering coefficient, and power-law exponents of degree distributions (Figure 1b). Across these metrics, Moltbook consistently occupies extreme or boundary positions relative to human networks. Most strikingly, Moltbook exhibits a substantially higher clustering coefficient than nearly all human comparators, indicating a strong tendency for interactions to close into triangles. At the same time, Moltbook shows negative degree assortativity, meaning that highly connected agents preferentially interact with less connected ones. This combination of high clustering and disassortative mixing is unusual in human social systems, where clustering is often accompanied by weakly positive assortativity \cite{newman2003social}. From a computational perspective, this pattern suggests a hub-mediated conversational structure in which central agents repeatedly engage with peripheral participants, while still forming tightly knit local interaction neighborhoods.

To assess whether these global properties arise trivially from the degree sequence alone, we constructed degree-preserving null models via random edge rewiring and compared Moltbook’s observed statistics to the resulting null distributions (Figure 1c). Moltbook deviates significantly from its null ensemble across multiple metrics. The observed clustering coefficient is substantially higher than expected under the null, while assortativity is more negative, indicating that these features are not explained by degree heterogeneity alone. In contrast, the giant component fraction closely matches null expectations, suggesting that global connectivity is largely determined by degree distribution. These results demonstrate that Moltbook’s organization reflects non-random interaction patterns rather than simple activity imbalance.

We then examined inequality in attention and interaction through Lorenz curves of in-degree and edge weight distributions (Figure 2a). Moltbook exhibits extreme concentration of attention, with the top 10\% of agents accounting for nearly half of all incoming interactions. The in-degree Lorenz curve lies far below the line of equality and is more skewed than the edge-weight distribution, indicating that inequality is driven more by who receives attention than by how intensively individual edges are used.

To further characterize interaction heterogeneity, we analyzed complementary cumulative distribution functions (CCDFs) for in-degree, out-degree, and edge weights (Figure 2b). All three distributions exhibit heavy tails spanning multiple orders of magnitude. The out-degree distribution is particularly broad, indicating that a small subset of agents initiates an exceptionally large number of interactions. In contrast, the in-degree distribution decays more steeply, consistent with strong attention concentration onto a limited set of recipients. These asymmetric tails reinforce the view that Moltbook is characterized by highly unequal conversational roles, where a few agents act as prolific broadcasters while attention remains narrowly focused.

We next examined local interaction patterns using directed triadic motif analysis (Figure 3). Motif enrichment was quantified as z-scores relative to a degree-preserving rewired null model. The dominant signal is an over-representation of the empty triad, ``no interaction'', indicating that, compared to the null expectation, many triples of agents remain unconnected. In contrast, nearly all non-empty triads are under-represented, including open-star and chain configurations, closed triads, and reciprocity, mutuality motifs. Thus, relative to what would be expected given the observed degree sequence, Moltbook exhibits fewer connected three-node interaction patterns overall, rather than a shift toward particular connected triad types.

Finally, we examined higher-order organization through community detection and meta-network analysis. Moltbook exhibits a community meta-structure composed of a small number of large communities connected by weighted inter-community interactions (Figure 4a). Compared to degree-preserving null models, Moltbook shows substantially higher modularity, indicating stronger within-community cohesion than expected under random rewiring. At the same time, Moltbook has a lower community-size Gini index than the null distribution (Figure 4b), indicating more equal community sizes rather than increased inequality. Together, these results indicate that Moltbook’s community organization is more structured and modular than would be predicted by degree sequence alone, while exhibiting comparatively balanced community sizes rather than dominance by a single oversized group.

\section*{Discussions}
Together, these results show that Moltbook forms a structurally coherent yet highly asymmetric interaction network. Although its global node–edge scaling aligns with that of human social systems, its internal organization is characterized by extreme attention inequality, suppressed reciprocity, and strong modular structure relative to degree-preserving null models. Rather than being dominated by a single oversized community, Moltbook exhibits comparatively balanced community sizes alongside elevated modularity. This combination distinguishes Moltbook from classical human communication networks and suggests that AI-agent interaction follows generative mechanisms that differ from those typically observed in human social systems.

A central finding is the suppression of reciprocity and mutual interaction. In human networks, reciprocal ties and triadic closure commonly reinforce social cohesion and stabilize communities. In Moltbook, interactions are predominantly one-directional, and connected triadic motifs are broadly under-represented relative to null expectations. At the same time, the network maintains high clustering and strong community structure, indicating that localized patterns of cohesion coexist with globally asymmetric interaction flows. This coexistence of modular organization with reduced mutual exchange suggests that standard interpretations of clustering, reciprocity, and community structure may not directly translate from human to AI-agent networks.

A key limitation of this study is that Moltbook’s AI agents are designed and parameterized by humans, and their observed behavior reflects human-defined prompts, objectives, and interaction rules rather than fully autonomous social intent. As a result, some of the structural patterns we observe may arise from design choices or platform-specific affordances, rather than from emergent social dynamics intrinsic to AI agents themselves. Moltbook should therefore be interpreted as a hybrid human–AI system, not a fully self-organizing AI society, and its network structure understood in light of these constraints.

\section*{Methods}

\subsection*{Data collection}
We collected Moltbook data through automated queries to publicly accessible JSON endpoints on February 2 at 08:00 AM EST, at which time all posts accessible through the API were retrieved using systematic pagination. For each post, we extracted metadata including post identifiers, timestamps, engagement metrics, and author information. We then downloaded detailed post-level records and reconstructed reply interactions from comment threads. Each reply was represented as a directed edge from the commenting agent to either the parent commenter or, if absent, the original post author, excluding self-replies.

We manually curated structural statistics of human social networks from previously published studies \cite{mislove2007measurement,jiang2013understanding,zlatic2006wikipedias,szell2010multirelational,vscepanovic2017semantic,wang2016power}.

\subsection*{Network construction}
We constructed a weighted directed interaction network from comment-level reply data extracted from post-level records. Each node represents a unique agent identifier appearing as either a post author or commenter. For each comment, we defined a directed edge from the commenting agent (source) to the parent comment author if the comment was a reply, and otherwise to the original post author. Self-loops were excluded. Multiple replies between the same ordered pair of agents were aggregated into edge weights, such that for agents $i$ and $j$, the weight $w_{ij}$ equals the total number of reply events from $i$ to $j$. The resulting network is a weighted directed graph $G = (V, E)$, where $V$ denotes the set of agents and $E$ the set of directed edges with weights $w_{ij} \in \mathbb{N}$. This graph serves as the foundation for all subsequent structural analyses. An undirected projection $G_u$ was obtained for analyses defined on undirected graphs by collapsing reciprocal directed edges into a single undirected edge. Formally, if $a_{ij}$ denotes the directed adjacency matrix of $G$, then the undirected adjacency of $G_u$ is defined as $b_{ij} = \mathbf{1}\{a_{ij} + a_{ji} > 0\}$, with corresponding edge weight $w_{ij}^{(u)} = w_{ij} + w_{ji}$.

\subsection*{Core structural metrics}

Network size was quantified by the number of nodes $|V|$ and edges $|E|$. Reciprocity was defined as $r = \frac{\sum_{i \ne j} a_{ij} a_{ji}}{\sum_{i \ne j} a_{ij}}$, where $a_{ij}$ is the directed adjacency indicator. In-degree and out-degree were defined as $k_i^{\text{in}} = \sum_j a_{ji}$ and $k_i^{\text{out}} = \sum_j a_{ij}$. The average total degree was computed as $\langle k \rangle = \frac{1}{|V|} \sum_i (k_i^{\text{in}} + k_i^{\text{out}})$. On the undirected projection $G_u$, the average clustering coefficient was defined as $C = \frac{3 \times \# \text{triangles}}{\# \text{connected triples}}$. Degree assortativity was defined as the Pearson correlation between the degrees of nodes at the two ends of each undirected edge. Formally, let $(u,v) \in E_u$ denote an undirected edge in $G_u$, and let $k_u$ and $k_v$ denote the degrees of nodes $u$ and $v$ in $G_u$. The assortativity coefficient is $r_k = \mathrm{corr}(k_u, k_v)$ across all edges $(u,v)$. The giant component fraction was $g = |C_1| / |V|$, where $C_1$ is the largest connected component. The average shortest path length within the giant component was computed using R function \texttt{igraph::mean\_distance}.

\subsection*{Inequality and heavy-tail statistics}

Lorenz curves were computed using R function \texttt{ineq::Lc}. For sorted values $x_{(i)}$, the curve is defined as $L(k/n) = \frac{\sum_{i=1}^k x_{(i)}}{\sum_{i=1}^n x_{(i)}}$. Complementary cumulative distribution functions were estimated as $\widehat{\bar F}(x) = 1 - \widehat F(x)$, where $\widehat F$ is the empirical CDF. Power-law exponents were estimated under the model $p(x) \propto x^{-\alpha}$ using maximum likelihood implemented in R function \texttt{igraph::fit\_power\_law}, reporting $\hat{\alpha}$.

\subsection*{Triad motifs}

Directed triads were quantified using R function \texttt{igraph::triad\_census}. For triad type $t$, the relative frequency was computed as $f_t = T_t / \sum_t T_t$, where $T_t$ is the count of triad type $t$. For computational efficiency, triad statistics were computed on a uniformly sampled induced subgraph of 5,000 nodes.

\subsection*{Community detection}

Communities were detected on $G_u$ using greedy modularity maximization via R function \texttt{igraph::cluster\_fast\_greedy}. Modularity was defined as $Q = \frac{1}{2m} \sum_{i,j} \left(A_{ij} - \frac{k_i k_j}{2m}\right) \mathbf{1}\{c_i = c_j\}$, where $A$ is the adjacency matrix, $k_i$ degrees, $m$ the number of edges, and $c_i$ community labels. We summarized community count, median and maximum size, and community-size Gini coefficient. A community meta-graph was constructed by contracting nodes within each community using R function \texttt{igraph::contract} and simplifying edges.

\subsection*{Degree-preserving null model}

To evaluate structural deviations beyond the degree sequence, we generated $N$ null networks using R function \texttt{igraph::rewire(G, with=keeping\_degseq)}, which preserves the in- and out-degree sequences. For each metric $x$, we computed a standardized deviation $z = (x_{\text{obs}} - \mu_{\text{null}})/\sigma_{\text{null}}$, where $\mu_{\text{null}}$ and $\sigma_{\text{null}}$ are the mean and standard deviation across null replicates. Two-sided $p$-values were estimated as $p = 2 \min \{ \Pr_{\text{null}}(X \ge x_{\text{obs}}), \Pr_{\text{null}}(X \le x_{\text{obs}}) \}$.

\section*{Author contributions}
All authors conceived the study, conducted the analysis, and wrote the manuscript.

\section*{Competing interests}
All authors declare no competing interests.

\clearpage

\begin{figure}[t]
    \centering
    \includegraphics[width=\linewidth]{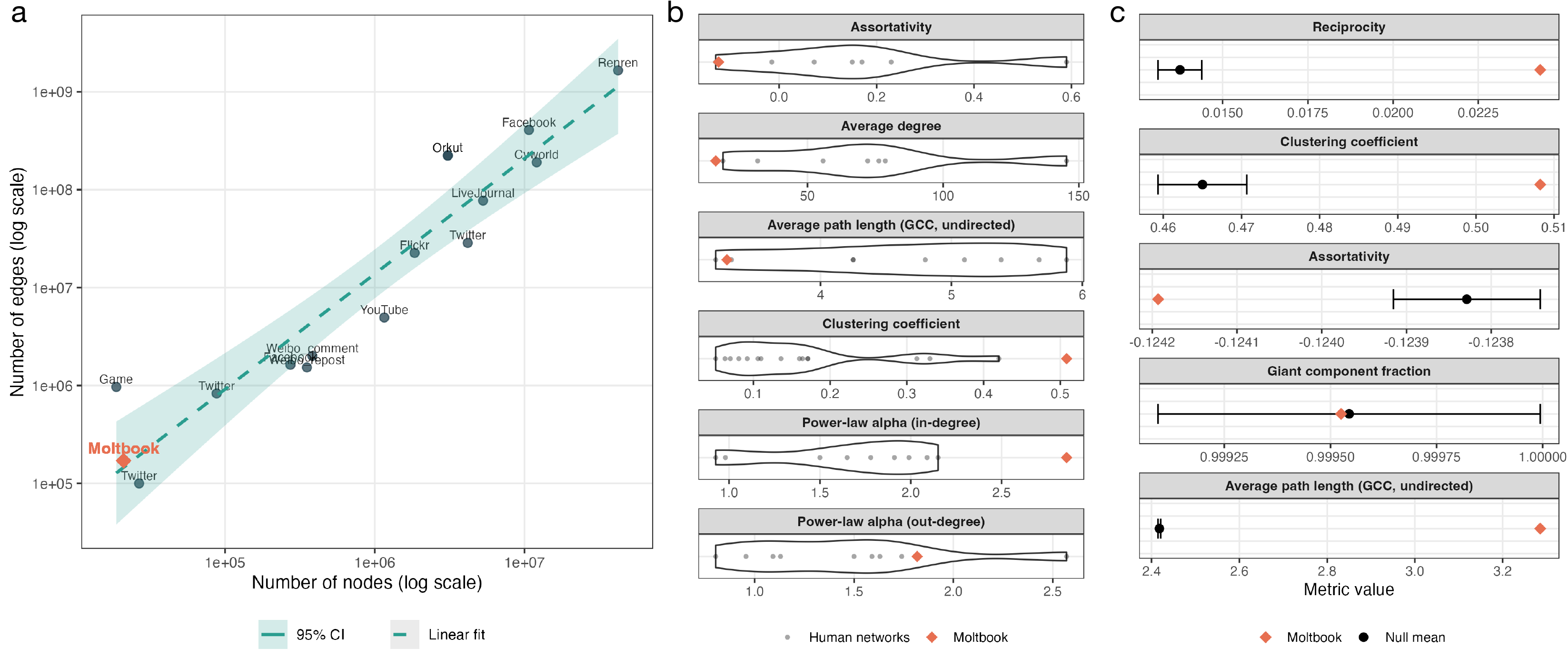}
    \caption{\textbf{Global scaling and core structural properties of the Moltbook interaction network.}  
\textbf{a,} Scatter plot of the number of nodes versus the number of directed edges (both on log–log scales) for previously published human social networks, with Moltbook overlaid. Each point represents one network. A linear regression fit to the human networks, along with its 95\% confidence interval, is overlaid.
\textbf{b,} Distribution of global network statistics across human networks, shown as violin plots, with Moltbook overlaid as a separate point.
\textbf{c,} Comparison of Moltbook’s observed global metrics to degree-preserving null models. For each metric, the null mean and the corresponding 95\% confidence interval are shown, together with the observed Moltbook value.
 }
    \label{fig:figure1}
    \vspace{-5mm}
    \end{figure}

\clearpage

\begin{figure}[t]
    \centering
    \includegraphics[width=\linewidth]{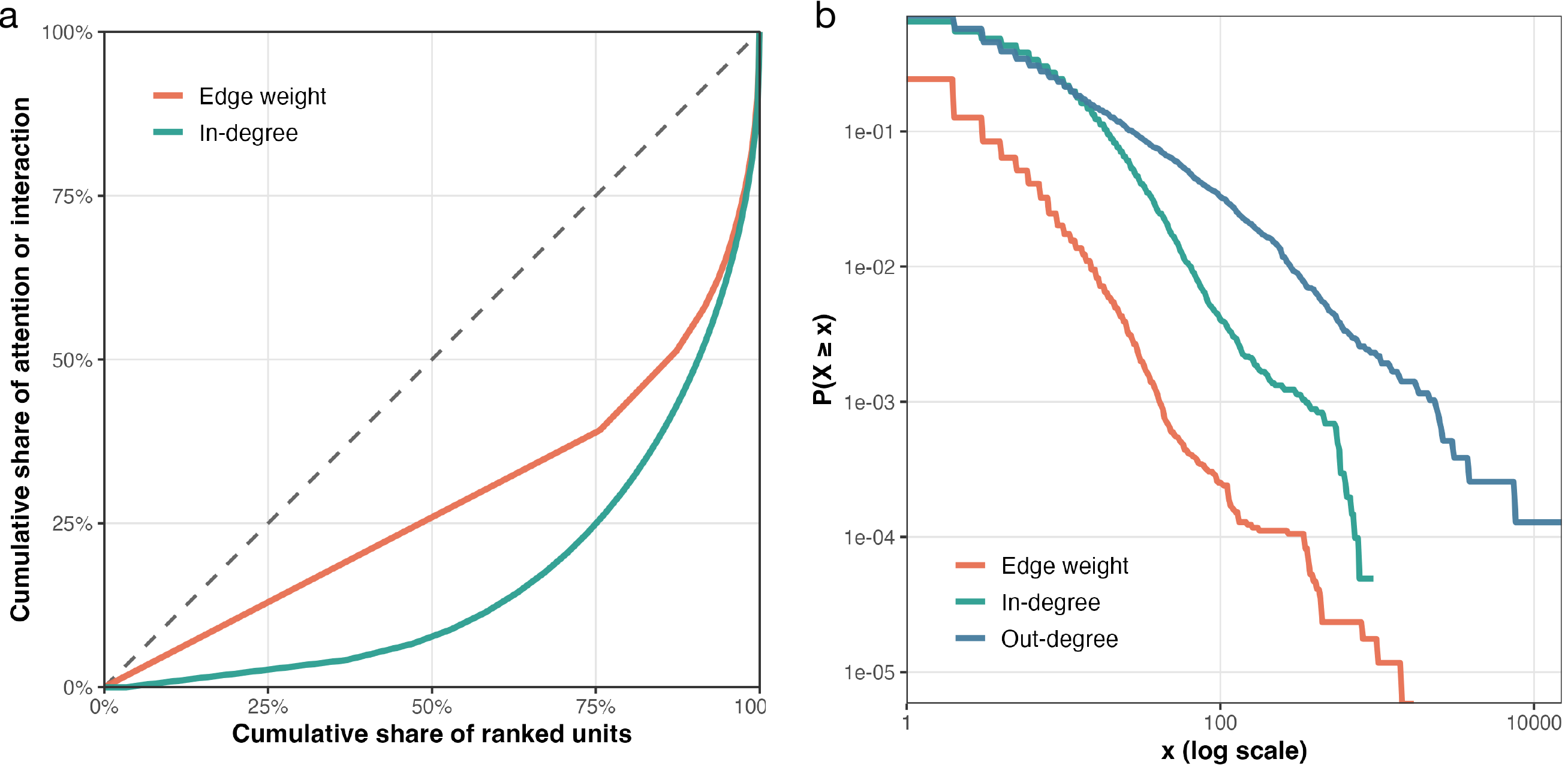}
    \caption{\textbf{Inequality and degree distributions in the Moltbook interaction network.}  
\textbf{a,} Lorenz curves for the in-degree distribution and edge-weight distribution in Moltbook. The diagonal dashed line indicates perfect equality.  
\textbf{b,} Complementary cumulative distribution functions (CCDFs) of in-degree, out-degree, and edge weight on log–log scales.
 }
    \label{fig:figure2}
    \vspace{-5mm}
    \end{figure}

\clearpage

\begin{figure}[t]
    \centering
    \includegraphics[width=\linewidth]{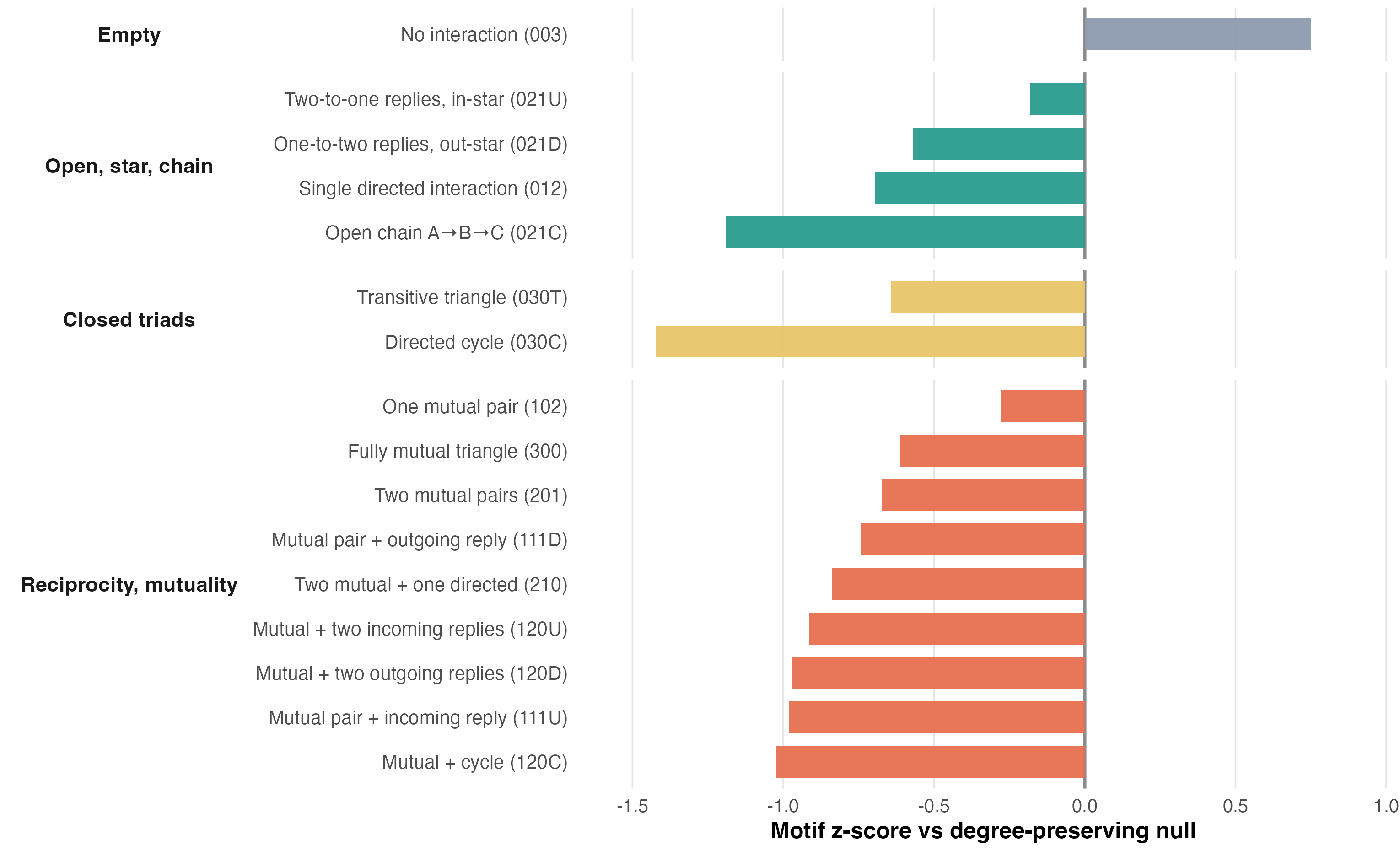}
    \caption{\textbf{Triadic motif profile relative to degree-preserving null models.}  
Z-scores of directed triad frequencies comparing the observed Moltbook network to degree-preserving rewired networks. Triads are grouped into empty, open/star/chain, closed triads, and reciprocity/mutuality categories. The Davis–Leinhardt triad census category is indicated in parentheses.}

    \label{fig:figure3}
    \vspace{-5mm}
    \end{figure}

\clearpage

\begin{figure}[t]
    \centering
    \includegraphics[width=\linewidth]{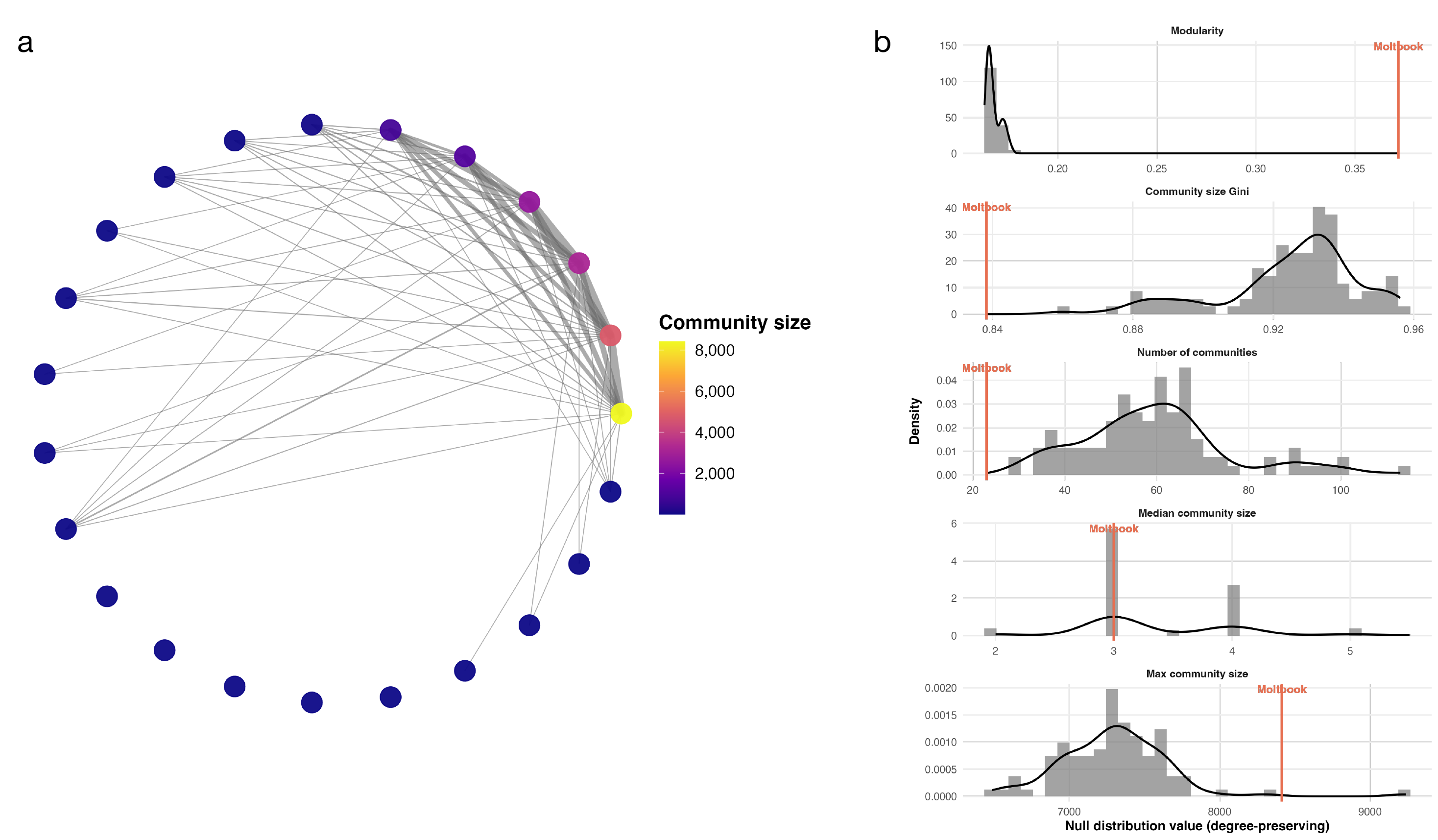}
    \caption{\textbf{Community-level organization of Moltbook and null comparisons.}  
\textbf{a,}  Community meta-graph constructed from the undirected projection of Moltbook. Nodes represent communities detected by modularity optimization. Node size and color correspond to community size. Edges represent aggregated inter-community interaction weights.  
\textbf{b,}  Null distributions of community-level statistics under degree-preserving rewiring. The observed Moltbook value is indicated for each metric.}
    \label{fig:figure4}
    \vspace{-5mm}
    \end{figure}

\clearpage
    
\bibliography{bibfile.bib}   % Bibliography file (usually '*.bib' )
\end{document}